\documentstyle[12pt]{article}

\catcode`\@=11
\long\def\@makefntext#1{ 
\protect\noindent \hbox to 3.2pt {\hskip-.9pt
$^{{\ninerm\@thefnmark}}$\hfil}#1\hfill} 

\def\thefootnote{\fnsymbol{footnote}}
 \def\@makefnmark{\hbox to 0pt{$^{\@thefnmark}$\hss}}  

\def\ps@myheadings{\let\@mkboth\@gobbletwo
\def\@oddhead{\hbox{} 
\rightmark\hfil\ninerm\thepage}
\def\@oddfoot{}\def\@evenhead{\ninerm\thepage\hfil 
\leftmark\hbox{}}\def\@evenfoot{}
\def\sectionmark##1{}\def\subsectionmark##1{}}

\def\slashmark#1#2#3{\global\setbox0=\hbox{\raise#2em
        \hbox{\kern#3em $#1\mathchar"0236$}}%
        \wd0=0pt \ht0=0pt \dp0=0pt \box0}
\def\dslash{{\mathchoice{\slashmark\displaystyle{.075}{-.075}}%
                {\slashmark\textstyle{.075}{-.075}}%
                {\slashmark\scriptstyle{.055}{-.055}}%
                {\slashmark\scriptscriptstyle{.04}{-.04}}\partial}}

\def\Lslash#1{{\mathchoice{\slashmark\displaystyle{-.1}{0}}%
                {\slashmark\textstyle{-.1}{-.125}}%
                {\slashmark\scriptstyle{-.075}{-.1}}%
                {\slashmark\scriptscriptstyle{-.06}{-.08}}#1}}

\begin{document}

\newcommand{\symbolfootnote}{\renewcommand{\thefootnote}
	{\fnsymbol{footnote}}}
\renewcommand{\thefootnote}{\fnsymbol{footnote}}
\newcommand{\alphfootnote}
	{\setcounter{footnote}{0}
	 \renewcommand{\thefootnote}{\sevenrm\alph{footnote}}}

\def\section{\@startsection{section}{1}
{\z@}{.5truecm plus -0.5ex minus -0.1ex}{\medskipamount}{\bf}}
\def\subsection{\@startsection{subsection}{2}
{\z@}{\bigskipamount}{\smallskipamount}{\it}}
\def\subsubsection{\@startsection{subsubsection}{3}
{\z@}{\bigskipamount}{\smallskipamount}{\rm}}


\def\thesection       {\arabic{section}.}
\def\thesubsection    {\thesection\arabic{subsection}.}
\def\thesubsubsection {\thesubsection \arabic{subsubsection}.}

\def\appendix{\par
  \setcounter{section}{0}
  \setcounter{subsection}{0}
  \def\thesection{\Alph{section}.}}

\def\Abstract#1{{
	\centering{\begin{minipage}{30pc}\tenrm\baselineskip=12pt\noindent
	\centerline{\tenrm ABSTRACT}\vspace{0.3cm}
	\parindent=0pt #1
	\end{minipage} }\par}}

\newcommand{\bibit}{\it}
\newcommand{\bibbf}{\bf}
\renewenvironment{thebibliography}[1]
	{\begin{list}{\arabic{enumi}.}
	{\usecounter{enumi}\setlength{\parsep}{0pt}
\setlength{\leftmargin 1.25cm}{\rightmargin 0pt}
	 \setlength{\itemsep}{0pt} \settowidth
	{\labelwidth}{#1.}\sloppy}}{\end{list}}

\topsep=0in\parsep=0in\itemsep=0in
\parindent=1.5pc

\newcounter{itemlistc}
\newcounter{romanlistc}
\newcounter{alphlistc}
\newcounter{arabiclistc}
\newenvironment{itemlist}
    	{\setcounter{itemlistc}{0}
	 \begin{list}{$\bullet$}
	{\usecounter{itemlistc}
	 \setlength{\parsep}{0pt}
	 \setlength{\itemsep}{0pt}}}{\end{list}}

\newenvironment{romanlist}
	{\setcounter{romanlistc}{0}
	 \begin{list}{$($\roman{romanlistc}$)$}
	{\usecounter{romanlistc}
	 \setlength{\parsep}{0pt}
	 \setlength{\itemsep}{0pt}}}{\end{list}}

\newenvironment{alphlist}
	{\setcounter{alphlistc}{0}
	 \begin{list}{$($\alph{alphlistc}$)$}
	{\usecounter{alphlistc}
	 \setlength{\parsep}{0pt}
	 \setlength{\itemsep}{0pt}}}{\end{list}}

\newenvironment{arabiclist}
	{\setcounter{arabiclistc}{0}
	 \begin{list}{\arabic{arabiclistc}}
	{\usecounter{arabiclistc}
	 \setlength{\parsep}{0pt}
	 \setlength{\itemsep}{0pt}}}{\end{list}}

%
%
%
%
%
\def\citen#1{%
\edef\@tempa{\@ignspaftercomma,#1, \@end, }
\edef\@tempa{\expandafter\@ignendcommas\@tempa\@end}%
\if@filesw \immediate \write \@auxout {\string \citation {\@tempa}}\fi
\@tempcntb\m@ne \let\@h@ld\relax \def\@citea{}%
\@for \@citeb:=\@tempa\do {\@cmpresscites}%
\@h@ld}
%
\def\@ignspaftercomma#1, {\ifx\@end#1\@empty\else
   #1,\expandafter\@ignspaftercomma\fi}
\def\@ignendcommas,#1,\@end{#1}
%
%
\def\@cmpresscites{%
 \expandafter\let \expandafter\@B@citeB \csname b@\@citeb \endcsname
 \ifx\@B@citeB\relax 
    \@h@ld\@citea\@tempcntb\m@ne{\bf ?}%
    \@warning {Citation `\@citeb ' on page \thepage \space undefined}%
 \else
    \@tempcnta\@tempcntb \advance\@tempcnta\@ne
    \setbox\z@\hbox\bgroup 
    \ifnum0<0\@B@citeB \relax
       \egroup \@tempcntb\@B@citeB \relax
       \else \egroup \@tempcntb\m@ne \fi
    \ifnum\@tempcnta=\@tempcntb 
       \ifx\@h@ld\relax 
          \edef \@h@ld{\@citea\@B@citeB }%
       \else 
          \edef\@h@ld{\hbox{--}\penalty\@highpenalty
            \@B@citeB }%
       \fi
    \else   
       \@h@ld\@citea\@B@citeB
       \let\@h@ld\relax
 \fi\fi%
 \def\@citea{,\penalty\@highpenalty\hskip.13em plus.1em minus.1em}%
}
%
%
\def\cite{\leavevmode\unskip\@ifnextchar[{\@tempswatrue\@citew}%
            {\@tempswafalse\@citex}}
%
%
\def\@citew[#1]#2{\ifnum\lastpenalty=\z@ \penalty\@highpenalty \fi
   \ [{\multiply\@highpenalty 3 
   \citen{#2}},\penalty\@highpenalty\ #1]\spacefactor\@m}
%
%
\def\@citex#1{\begingroup \leavevmode \@tempcnta\@m \unskip
  \def\@tempa{\@cite{\citen{#1}}\spacefactor\@tempcnta\endgroup}%
  \futurelet\@tempb\@citey}%
%
%
\def\@citey{\let\@tempc\@tempa
   \ifx\@tempb.\ifnum\spacefactor>2999 \let\@tempb\relax\fi\let\@tempc\@citez
   \else\ifx\@tempb,\let\@tempc\@citez
   \else\ifx\@tempb:\let\@tempc\@citez 
   \else\ifx\@tempb;\let\@tempc\@citez 
   \fi\fi\fi\fi
   \@tempc}%
%
\def\@citez#1{\@tempcnta\sfcode`#1\@tempb\futurelet\@tempb\@citey}%
\def\@cite#1{$\m@th\the\scriptfont\z@\edef\bf{\the\scriptfont\bffam}%
      ^{\hbox{#1}}$}
\let\nocitecount\relax 
\def\fnm#1{$^{\mbox{\scriptsize #1}}$}
\def\fnt#1#2{\footnotetext{\kern-.3em
	{$^{\mbox{\sevenrm #1}}$}{#2}}}

\font\twelvebf=cmbx10 scaled\magstep 1
\font\twelverm=cmr10 scaled\magstep 1
\font\twelveit=cmti10 scaled\magstep 1
\font\elevenbfit=cmbxti10 scaled\magstephalf
\font\elevenbf=cmbx10 scaled\magstephalf
\font\elevenrm=cmr10 scaled\magstephalf
\font\elevenit=cmti10 scaled\magstephalf
\font\bfit=cmbxti10
\font\tenbf=cmbx10
\font\tenrm=cmr10
\font\tenit=cmti10
\font\ninebf=cmbx9
\font\ninerm=cmr9
\font\nineit=cmti9
\font\eightbf=cmbx8
\font\eightrm=cmr8
\font\eightit=cmti8

{\normalsize \hfill UW/PT-94-10}
\vspace{1cm}
\thispagestyle{empty}

\centerline{\bf MOTION OF SPIN-$1/2$ PARTICLES IN EXTERNAL}
\vspace{0.4cm}
\centerline{\bf GRAVITATIONAL AND ELECTROMAGNETIC FIELDS}
\vspace{2cm}

\centerline{{\sc Stamatis Vokos}\footnote{E-Mail: vokos@phys.washington.edu}}
\baselineskip=15pt
\centerline{\tenit Physics Department, FM-15}
\baselineskip=12pt
\centerline{\tenit University of Washington}
\baselineskip=12pt
\centerline{\tenit Seattle, WA 98195}
\baselineskip=12pt
\centerline{\tenit USA}
\vspace{0.9cm}

\renewcommand{\arraystretch}{2.0}
\renewcommand{\thefootnote}{\alph{footnote}}
\newcommand{\beq}{\begin{equation}}
\newcommand{\eeq}{\end{equation}}
\newcommand{\beqa}{\begin{eqnarray}}
\newcommand{\eeqa}{\end{eqnarray}}
\vspace{10mm}

\Abstract{
We study the semi-classical limit of the solution of the Dirac
equation in a background electromagnetic/gravitational plane wave.
We show that the exact solution corresponding to an asymptotically
fixed incoming momentum satisfies constraints consistent with the
classical notion of a spinning particle. In order to further analyze
the motion of a spinning particle in this external
inhomogeneous field one has to consider wave-packet superpositions
of these exact solutions. We are currently investigating
the existence of a {\tenit classical\/} theory of a phenomenological
spin tensor which reproduces our quantum-mechanical results.
}
\vspace*{10mm}

\vfill
\baselineskip=14pt

\begin{center}
To appear in the proceedings of the\\
{\sl 7th Marcel Grossmann Meeting} \\
{\sl on General Relativity (MG 7) } \\
Stanford, CA, July 24  -- 30, 1994
\end{center}

\vfill

\newpage

\twelverm   
\baselineskip=14pt

\section{Introduction}

There has been a resurgence of interest lately in the study of the
dynamics of a spinning object in an external field. Indeed, there
are special cases where the answer is well-known, examples of which
are the case of a microscopic particle in a (nearly) uniform
electromagnetic field and the formal description of a macroscopic
spin in a gravitational field with special symmetries. The former
case is described by the Michel-Bargmann-Telegdi (BMT) equation,
while the latter is formally solved within the Papapetrou-Dixon
theoretical framework. One may argue that the general case is
well-defined, since the coupling of spinning matter with external
fields is known to be through the appropriate field equation in the
given background field. This approach, however, is lacking in
practicality. The exact solutions of any field equation in a background field
are very few and most do not correspond to physically
realistic configurations.  Clearly, one does not resort to solving the Dirac
equation in order to predict the behavior of an electron spin in an
accelerator (cf. Siberian snakes)! One is then drawn to the
challenging task of describing this interaction of spinning particles in
terms of a set of self-consistent equations for the components of a
phenomenological spin tensor which satisfy certain constraints. The
approaches to this problem have been varied (see Ref.~\citen{vokos}
and references therein). Researchers have employed arguments
of analogy with the BMT equation, have used Lagrangian theories, Hamiltonian
constructions, Ruthian formulations, supersymmetric manipulations,
quaternion and octonion algebras, etc.

In this talk, I motivate the use of a specific toy model where
one may be able to test the extant theories of spin-${1\over 2}$
particles in external fields.
My philosophical bias is that {\em no} (semi-)classical theory
of spinning particles in external fields can ever hope to be
acceptable if it does not
agree with a (properly chosen) classical limit of the corresponding
field equation. It is useful, therefore, to start with the Dirac
equation and construct a suitable phenomenological spin tensor, whose time
development is governed by equations deduced from
the quantum mechanical dynamics.

\section{The model}

There exists a particularly simple (but non-trivial),
realistic gravitational and electromagnetic background, where one can
solve the Dirac equation exactly, namely in a plane gravitational or
electromagnetic wave. The wave's spectral decomposition can be arbitrary,
guaranteeing departure from uniformity. The high degree of symmetry,
however, provides one with a wonderfully tractable solution. The
solution in the electromagnetic case is originally due to Volkov. Here
we lean heavily on the description of Brown\cite{lsb}. Due to the formal
similarity of the electromagnetic and gravitational backgrounds, we were
able to easily obtain a solution for the gravitational case, as well.
In the interest of brevity, we present here the outline of the
argument in the electromagnetic case only.

A plane wave is given by a vector potential $A_\mu$ which
depends on a single argument only, i.e.
\begin{equation}
A_\mu = A_\mu (y)\,,
\end{equation}
where $y=n_\mu x^\mu$, and $n^\mu$ is a null vector which describes
the unique direction of propagation of the plane wave. In the Lorentz
gauge $\partial\cdot A=0=n\cdot A$, one can obtain the solution of the
Dirac equation
\begin{equation}
(i\dslash -e \Lslash{A} -m)\psi_p(x)=0\,,
\end{equation}
where the subscript $p$ in the solution refers to the asymptotic incoming
momentum of the fermion.  The explicit form of the
solution\footnote{
One interesting point to note here, however, is that one can arrive at
the solution in the following way. One solves the classical Lorentz
force equation (or geodesic equation, in the gravitational case), in
terms of Hamilton-Jacobi theory, and one obtains explicitly Hamilton's
characteristic function, $\phi$. Armed with $\phi$, one can show
that the exponential of $\phi$, $\exp(i\phi/\hbar)$, satisfies the
Klein-Gordon equation in the relevant background field, i.e.\ the WKB
approximation is exact in this case. The final step is to verify that
the solution to the Dirac equation in the same background is
equal to the WKB factor multiplied by an appropriate spinor. The beauty
of this approach, is that one is always in touch with the classical
interpretation of quantum mechanical observables.  One example is
the establishment of the proportionality between $y=n\cdot x$ and the
classical proper time $\tau$, $y\propto \tau$.} together
with the accompanying details are given elsewhere\cite{vokos,lsb}.

One can perform a Gordon decomposition of the Dirac current,
$j^\mu_p(x)$,
\beqa
j^\mu_p(x)&=&\overline{\psi_p}(x)\gamma^\mu\psi_p(x)\equiv
              j_{\rm\bf c}\,^\mu(x)+j_{\rm\bf s}\,^\mu(x)\nonumber\\
          &=&{i\hbar\over 2m}\left(\overline{\psi_p}(x)D^\mu\psi_p(x)-
                             D^\mu\overline{\psi_p}(x)\psi_p(x)\right)+
             {\hbar\over 2m}
      \partial_\nu\left(\overline{\psi_p}(x)\sigma^{\mu\nu}\psi_p(x)\right)\,.
\eeqa
The first term in the above decomposition is the convective current and the
second is the spin current. Both currents are separately conserved.
One obtains,
\beqa
j_{\rm\bf c}\,^\mu(y) &=& {1\over m}\left(p^\mu - e A^\mu(y) + n^\mu
I_p(y)\right)+ {e\hbar\over 4m}{n^\mu\over n\cdot p}\overline{u}(p)
F_{\alpha\beta}(y)\sigma^{\alpha\beta} u(p)\\
j_{\rm\bf s}\,^\mu(y) &=& -{e\hbar\over 4m}{n^\mu\over n\cdot p}\overline{u}(p)
F_{\alpha\beta}\sigma^{\alpha\beta} u(p)
\equiv {1\over m} \partial_\nu J^{\mu\nu}\,,
\eeqa
where $I_q(y)={1\over 2n\cdot q} (2 e A\cdot q - e^2 A^2)$ and $u(p)$
is a free Dirac spinor. Note that each current depends only on $y$,
and that the total current is independent of the spin, through a
non-trivial cancellation of terms.\footnote{
This shows the error in the usual
statement that the convective current is the part of the current which
corresponds to the translational motion only and is, therefore,
independent of spin.}

A natural candidate for a semi-classical spin tensor is ${\cal
S}^{\mu\nu}$, where
\beq
J^{\mu\nu}(y)= \overline{\psi_p}(x){\hbar\over 2}\sigma^{\mu\nu}\psi_p(x)
\equiv \overline{u}(p){\cal S}^{\mu\nu}(y)u(p)\,.
\eeq
Exploiting the classical equality $y= {n\cdot p\over m} \tau\,,$
we are then able to express the spin tensor of the fermion at proper time
$\tau$ in terms of the initial spin tensor at $\tau\to -\infty$,
\beq
{\cal S}^{\mu\nu}(\tau)={\cal O}^{\mu\nu}\,_{\rho\lambda}(\tau){\cal
S}^{\rho\lambda}(-\infty)\,,
\eeq
where
\beqa
{\cal O}^{\mu\nu}\,_{\rho\lambda}(\tau)=
\delta^\mu_\rho \delta^\nu_\lambda \! &+& \!
{2 e\over m}\int_{-\infty}^\tau\!\! d\tau'\,
         \delta_\rho\,^{[\mu} F^{\nu]}\,_\lambda(\tau')+
{e^2\over 2m^2}\int_{-\infty}^\tau \!\!\! d\tau'\int_{-\infty}^\tau \!\!\!
         d\tau''\,  F^{\mu\nu}(\tau')F_{\rho\lambda}(\tau'')\nonumber \\
 &+&
{e^2\over(n\cdot p)^2}n^{[\mu}\delta^{\nu]}\,_\rho n_\lambda A^2(\tau)\,.
\eeqa
Note that the above expression for the spin tensor involves the vector
potential in the Lorentz
gauge. However, $\cal S$ is invariant under the restricted gauge transformation
$A_\mu(y)\to A_\mu(y) + \partial_\mu \chi(y)$, as one may easily verify.

This completely defines the dynamics of the spin tensor
in this specific background field. For additional details and for the
solution in the gravitational wave background see Ref.~\citen{vokos}.
Let us turn next to the issue of constraints. Any set of
dynamical equations for the spin tensor is incomplete
without conditions which restrict the number of components of the spin
tensor\cite{papapetrou,dixon}. In this case, one can show that
$\cal S$ satisfies the following constraints without the use of equations
of motion, namely
\beq
v_\mu(\tau) {\cal S}^{\mu\nu}(\tau) \equiv {1\over m} (p_\mu-e A_\mu +
n_\mu I_p(\tau)) {\cal S}^{\mu\nu}(\tau) = 0\,
\eeq
assuming $v_\mu(\tau\to -\infty) {\cal S}^{\mu\nu}(\tau\to -\infty) =
p_\mu/m \,\,{\cal S}^{\mu\nu}(\tau\to -\infty)= 0$, and
\beq
{d\over d\tau}\left({\cal S}^{\mu\nu}(\tau){\cal
S}_{\mu\nu}(\tau)\right)=0\,.
\eeq

\section*{Acknowledgements}
I am indebted to David Boulware, Lowell Brown, and Laurence Yaffe for
many useful discussions. This work was supported in part by DOE grant
DE-FG09-91ER40614 .

\end{document}